\documentclass[12pt,preprint]{aastex}

\newcommand{\myemail}{rfosbury@eso.org}
\newcommand{\kms}{km~s$^{-1}$}
\newcommand{\cmmt}{cm$^{-2}$}
\newcommand{\hii}{\ion{H}{2}}
\newcommand{\hi}{\ion{H}{1}}
\newcommand{\nv}{\ion{N}{5}}
\newcommand{\niv}{\ion{N}{4}]}
\newcommand{\niii}{\ion{N}{3}]}
\newcommand{\oiii}{\ion{O}{3}]}
\newcommand{\neb}{[\ion{O}{3}]}
\newcommand{\oii}{[\ion{O}{2}]}
\newcommand{\ciii}{\ion{C}{3}]}
\newcommand{\civ}{\ion{C}{4}}
\newcommand{\heii}{\ion{He}{2}}
\newcommand{\lya}{Ly$\alpha$}
\newcommand{\hb}{H$\beta$}
\newcommand{\neiii}{[\ion{Ne}{3}]}
\newcommand{\siiii}{\ion{Si}{3}]}
\newcommand{\siiv}{\ion{Si}{4}}
\newcommand{\ovi}{\ion{O}{6}}
\newcommand{\cii}{\ion{C}{2}}
\newcommand{\neiv}{[\ion{Ne}{4}]}
\newcommand{\mgii}{\ion{Mg}{2}}
\newcommand{\hei}{\ion{He}{1}}
\newcommand{\oi}{[\ion{O}{1}]}
\newcommand{\ha}{H$\alpha$}
\newcommand{\nii}{[\ion{N}{2}]}
\newcommand{\sii}{[\ion{S}{2}]}
\newcommand{\ujy}{$\mu$Jy}
\newcommand{\ecs}{erg cm$^{-2}$\ s$^{-1}$}
\newcommand{\ecsa}{erg cm$^{-2}$\ s$^{-1}$\AA$^{-1}$}
\newcommand{\object}{RX~J0848$+$4456}
\newcommand{\objectS}{CL~J0848.8$+$4455}
\newcommand{\map}{{\mbox{\sc mappings~i}c}}

\shorttitle{lensed \hii\ galaxy at $z=3.357$}
\shortauthors{Fosbury et al.}

\begin{document}


\title{MASSIVE STAR FORMATION IN A GRAVITATIONALLY-LENSED  \hii\ GALAXY 
         AT  z = 3.357\altaffilmark{1, 2, 3}}
        
\altaffiltext{1}{Based partly on observations with the NASA/ESA Hubble Space
Telescope, obtained at the Space Telescope Science Institute, which is
operated by the Association of Universities for Research in Astronomy,
Inc. under NASA contract No. NAS5-26555.}
\altaffiltext{2}{Some of the data presented herein were obtained at the W.M. Keck Observatory, which is operated as a scientific partnership among the California Institute of Technology, the University of California, and the National Aeronautics and Space Administration (NASA). The Observatory was made possible by the generous financial support of the W.M. Keck Foundation.}
\altaffiltext{3}{Based partly on observations obtained by staff of the Gemini Observatory, which is operated by the Association of Universities for Research in Astronomy, Inc., under a cooperative agreement with the NSF on behalf of the Gemini partnership: The National Science Foundation (United States), the Particle Physics and Astronomy Research Council (United Kingdom), the National Research Council (Canada), CONICYT (Chile), the Australian Research Council (Australia), CNPq (Brazil), and CONICET (Argentina). }


\author{R.A.E. Fosbury\altaffilmark{4}}
\affil{ST-ECF, Karl Schwarzschild Str. 2, D-85748, Garching, Germany}
\email{\myemail}

\author{M. Villar-Mart\'{\i}n, A. Humphrey}
\affil{University of Hertfordshire, Hatfield, Herts AL10 9AB, UK}

\author{M. Lombardi, P. Rosati}
\affil{ESO, D-85748, Garching, Germany}

\author{D. Stern}
\affil{Institute of Geophysics and Planetary Physics, Lawrence Livermore National Laboratory, Livermore, CA 94550, USA}

\author{R. N Hook\altaffilmark{5}}
\affil{Space Telescope Science Institute, Baltimore, MD 21218, USA}

\author{B. P. Holden, S. A. Stanford}
\affil{Department of Physics, University of California, Davis, CA 95616, USA}

\author{G. K. Squires}
\affil{SIRTF Science Center, California Institute of Technology, Pasadena, CA 91125, USA}

\author{M. Rauch}
\affil{The Observatories of the Carnegie Institution of Washington, 813 Santa Barbara Street, Pasadena, CA 91101, USA}

\and

\author{W.L.W. Sargent}
\affil{California Institute of Technology, Pasadena, CA 91125, USA}


\altaffiltext{4}{Affiliated to the Space Telescopes Division, Research and Space Science Department, European Space Agency.}

\altaffiltext{5}{Seconded from: ST-ECF, Karl Schwarzschild Str. 2, D-85748, Garching, Germany}


\begin{abstract}
The Lynx arc, with a redshift of 3.357, was discovered during
spectroscopic follow-up of the $z = 0.570$ cluster \object\ from
the ROSAT Deep Cluster Survey.  The arc is characterized by a
very red $R - K$ color and strong, narrow emission lines.  Analysis of
HST WFPC~2 imaging and Keck optical and infrared spectroscopy
shows that the arc is an \hii\ galaxy magnified by a factor
of $\sim 10$ by a complex cluster environment.  The high intrinsic
luminosity, the emission line spectrum, the absorption components seen
in Ly$\alpha$ and \civ, and the restframe ultraviolet continuum
are all consistent with a simple \hii\ region model containing $\sim 10^6$\ hot O stars.  The best fit
parameters for this model imply a very hot ionizing continuum ($T_{\rm
BB} \simeq 80,000$~K), high ionization parameter ($\log U \simeq -1$),
and low nebular metallicity ($Z / Z_\odot \simeq 0.05$).  The narrowness
of the emission lines requires a low mass-to-light ratio for the
ionizing stars, suggestive of an extremely low metallicity stellar cluster.  The apparent overabundance of silicon in the
nebula could indicate enrichment by past pair instability supernov\ae,
requiring stars more massive than $\sim140 M_\odot$.

\end{abstract}




\section{INTRODUCTION}
During a multiwavelength study of the cluster \object\ ($z = 0.570$) and its slightly lower redshift (0.543) companion, which we shall call \objectS,  from the ROSAT Deep Cluster Survey, \citet{hol01} discovered a gravitational arc at  $\alpha=$~08$^{\rm h}$48$^{\rm m}$48$^{\rm s}$.76; $\delta =+$44$\degr$55$'$49$''$.6 (J2000). This shows an unusual narrow emission line spectrum identified with ultraviolet H, He, C, N and O lines at a redshift of $z = 3.36$. Using arguments based on photoionization modelling, these authors suggested that the emission arises from low metallicity gas ionized, at a high ionization parameter ($U$ = photon/matter density), by a source with a blackbody temperature of between 80,000 and 100,000K. The color temperature of the ionizing source could be constrained with some confidence due to the dominant strength of the \niv\  intercombination doublet over the \niii\ and \nv\ lines in the observed wavelength range.

The two components of the arc and the extended, faint morphology suggest that the source lies along a fold caustic and so could be highly magnified. In this paper, we exploit the opportunity offered by this amplification to observe, in unprecedented detail, what appears to be a young, low metallicity star-forming region at an early epoch. In addition to the restframe UV spectra previously reported \citep{hol01}\ we present new low and intermediate ($R\approx 5000$) resolution optical spectra and H- and K-band spectra taken with the Keck telescopes. Together, these spectra allow the measurement of almost all the  emission lines expected to be reasonably strong from \lya\ to \neb$\lambda\lambda$4959, 5007. The line strengths and profiles are used to study the physical, chemical and kinematic properties of the gas and the ionizing stars and make comparisons with lower redshift \hii\ regions and starforming galaxies. HST and groundbased imaging data are used both for photometric calibration purposes and as a basis for the gravitational lensing model that allows us to estimate the magnification. This source, while at a somewhat lower redshift, may be similar in many respects to the very faint, high redshift \lya\ sources being found in the vicinity of clusters \citep{eli01,huc02} and in the field \citep{rho03, kod03}.

\section{OBSERVATIONS}

\subsection{Imaging}

Two sets of mosaics covering the cluster were obtained with the HST WFPC~2 (PI Rosati; proposal ID 7374) in filter F702W. The region containing the arc is within both sets of images and comprises a total exposure of 20 $\times$ 1200s \citep{hol01}.  An additional 4800s of exposure in filter F814W (PI Elston; proposal ID 8269) were also examined. 

All the images were retrieved from the ST-ECF archive and standard `on-the-fly' pipeline processing performed using the latest calibration pipeline and reference files. The resultant calibrated single images were still dithered relative to each other, geometrically distorted and contained numerous cosmic rays. These frames were subsequently combined using the `metadrizzle' script (Fruchter 2002, priv. comm.) which implements the scheme described in \citet{frh02} to both detect and flag the cosmic rays and other defects and perform an optimum combination of the shifted frames using the `drizzle' method. The other sets of images, taken at different times and hence having different pointings and orientations, were aligned using their image header world-coordinate information. Small final offsets to bring them into precise registration were determined from a study of objects in the overlap regions.

Given the short color baseline, the data from these two filters give little color information. The F702W data are shown in figure~\ref{hst-image}. Together, these filters are dominated by nebular continuum emission (see below) but also include several strong emission lines, notably \niv\ $\lambda\lambda$1483, 1486; \civ\ $\lambda\lambda$1548, 1550; \oiii\ $\lambda\lambda$1661, 1666 and \ciii\ $\lambda\lambda$1907, 1909.

\clearpage

\placefigure{hst-image}

\begin{figure}
\plotone{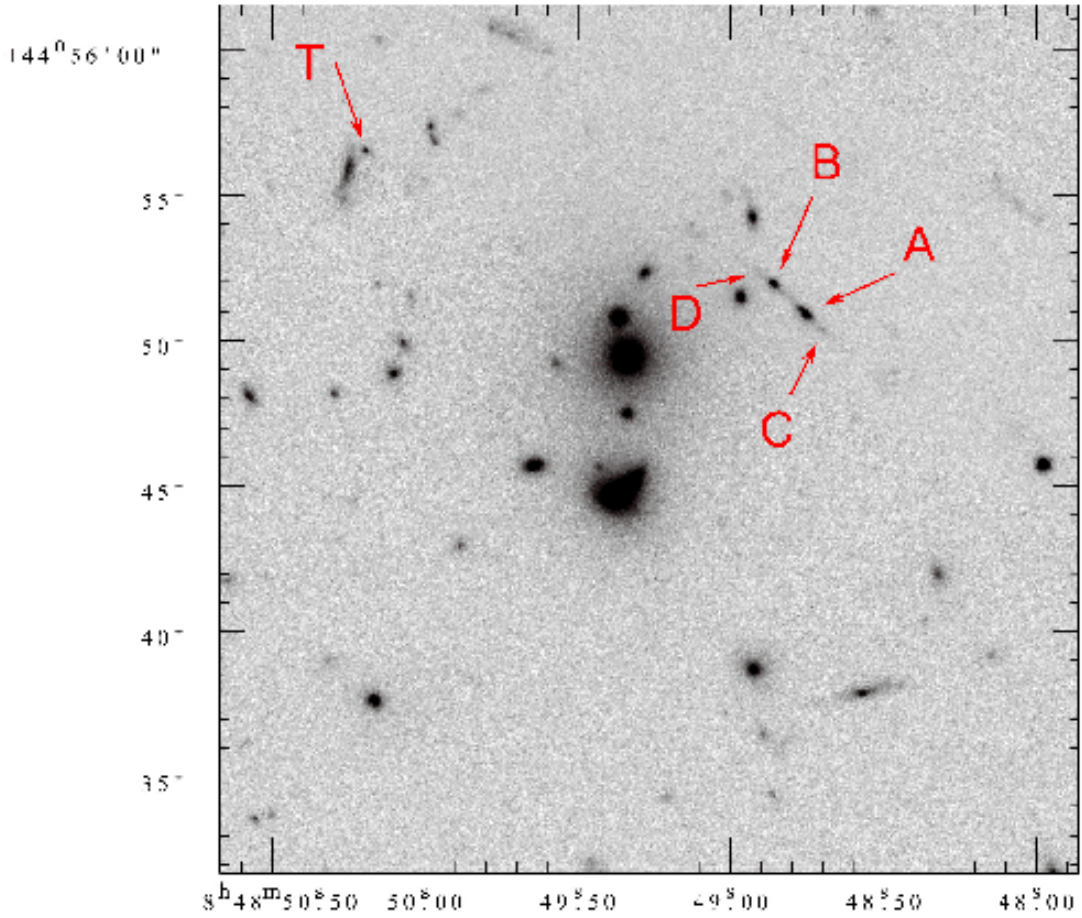}
\caption{A 30 arcsec square subset of the  F702W  image of the arc in Lynx from the HST WFPC~2
  camera. This is centred approximately on the southern cluster \objectS\ and is scaled to show the brightest cluster galaxies and the details of the gravitationally-lensed structures. The arc is labelled with its two bright components A and B. A thin arc extends over about $9''$\ with two very faint
  components, C and D, as well as other fainter substructure lying
  along it. A candidate third image, T, is discussed in
  Section~\ref{lensing}.\label{hst-image}}
\end{figure}

\clearpage

\subsection{Spectroscopy}

\begin{figure}
\plotone{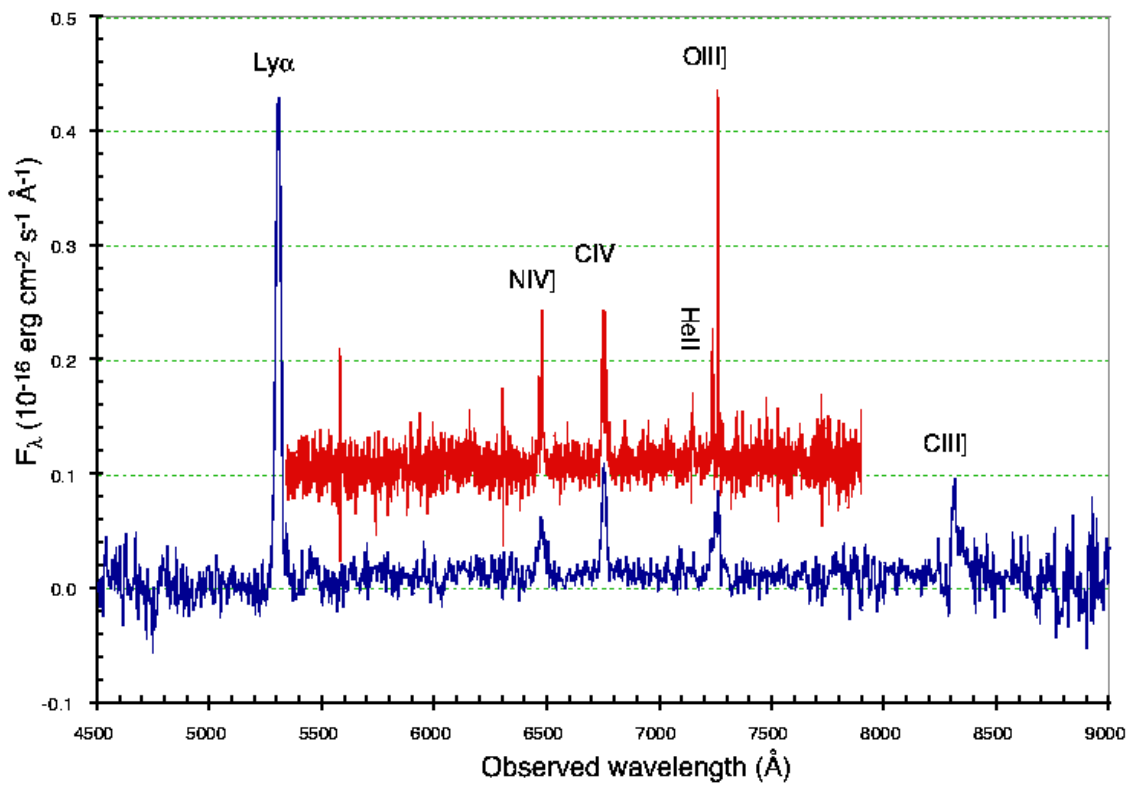}
\caption{The two LRIS spectra of the arc. The higher resolution spectrum has been shifted by 0.1 units of the flux scale ($10^{-16}$~\ecsa). Both spectra have been scaled by the photometric correction factors described in the text and so represent the sum of components A and B.\label{lris-spectra}}
\end{figure}

The log of spectroscopic observations is given in
table~\ref{spectroscopic-log}. The first spectrum, obtained with the
Keck~I telescope using the Low Resolution Imaging Spectrograph, LRIS,
\citep{oke95} is described in \citet{hol01}.  This (LRIS$_{\rm med}$) used
the 600 line mm$^{-1}$\ grating blazed at 7500\AA\ and was taken from a
multi-object mask with the slit lying along the arc, thus including
both the A and B components. The second, lower resolution LRIS
spectrum (LRIS$_{l\rm ow}$) was taken as part of the SPICES survey
\citep{ste02, eisip} and used the 150 line mm$^{-1}$ grating blazed at
7500\AA. The slit, again from a multi-object mask, was placed to include
the brightest arc component (A). The spectra were reduced and the data
extracted using standard scripts, optimised for LRIS, as described in
\citep{hol01}. One-dimensional extractions are shown in
figure~\ref{lris-spectra}.

\placetable{spectroscopic-log}

\placefigure{lris-spectra}

\clearpage

\begin{deluxetable}{lccccc}
\tablecaption{Log of Keck spectroscopic observations. \label{spectroscopic-log}}
\tablewidth{0pt}
\tablehead{
\colhead{UT date} & \colhead{Instrument}   & \colhead{$\lambda$-range}   &
\colhead{Dispersion} & \colhead{Slit/PA} & \colhead{Exposure}\\
&  & \colhead{\AA} & \colhead{km s$^{-1}$/pixel}  &\colhead{arcsec/deg} & \colhead{s}
}
\startdata
2000 Feb 3 & LRIS & 4500$-$9000  &55&$16 \times 1.5 / 45$ &5400\\
2000 Dec 23 & LRIS & 5340$-$7900& 220  & $20 \times 1.5$ / 202.5 &1800\\
2000 Mar 3 & ESI & 3900$-$11,000  & 11.4&$20 \times 1 $&3600\\
2001 Nov 20 & NIRSPEC & H \& K  &50& $42 \times 0.76 / 46.5$& 450$+$450\\
 \enddata


\tablecomments{The spectral resolution is typically $\sim3\times$ the dispersion.}


\end{deluxetable}

\clearpage

\clearpage

In order to measure the emission line profiles and doublet ratios, a
higher resolution spectrum, with a resolving power of $R \sim 4300$, was
obtained with the Echellette Spectrograph and Imager (ESI;
\citet{epp98} on the Keck~II telescope on the night of UT 2000 March
3. The slit was centred on the fainter of the two arc components (B in
figure~\ref{hst-image}).  These data were reduced and wavelength
calibrated using the MAuna Kea Echelle Extraction (MAKEE) package
developed by T. Barlow.  The main steps that this package performs are
(i) debias, (ii) fit order traces using the standard star image, (iii)
flat-field, (iv) mask cosmic rays and bad pixels, (v) extract spectrum
for each order and subtract night-sky spectrum, (vi) wavelength
calibrate for vacuum wavelengths.  Flux calibration was performed
using observations of the standard star G191B2B, correcting for
atmospheric extinction using the Mauna Kea extinction curve. ESI
spectra of \lya\ and \civ\ are presented in
section~\ref{line-spectrum}.

\begin{figure}
\plotone{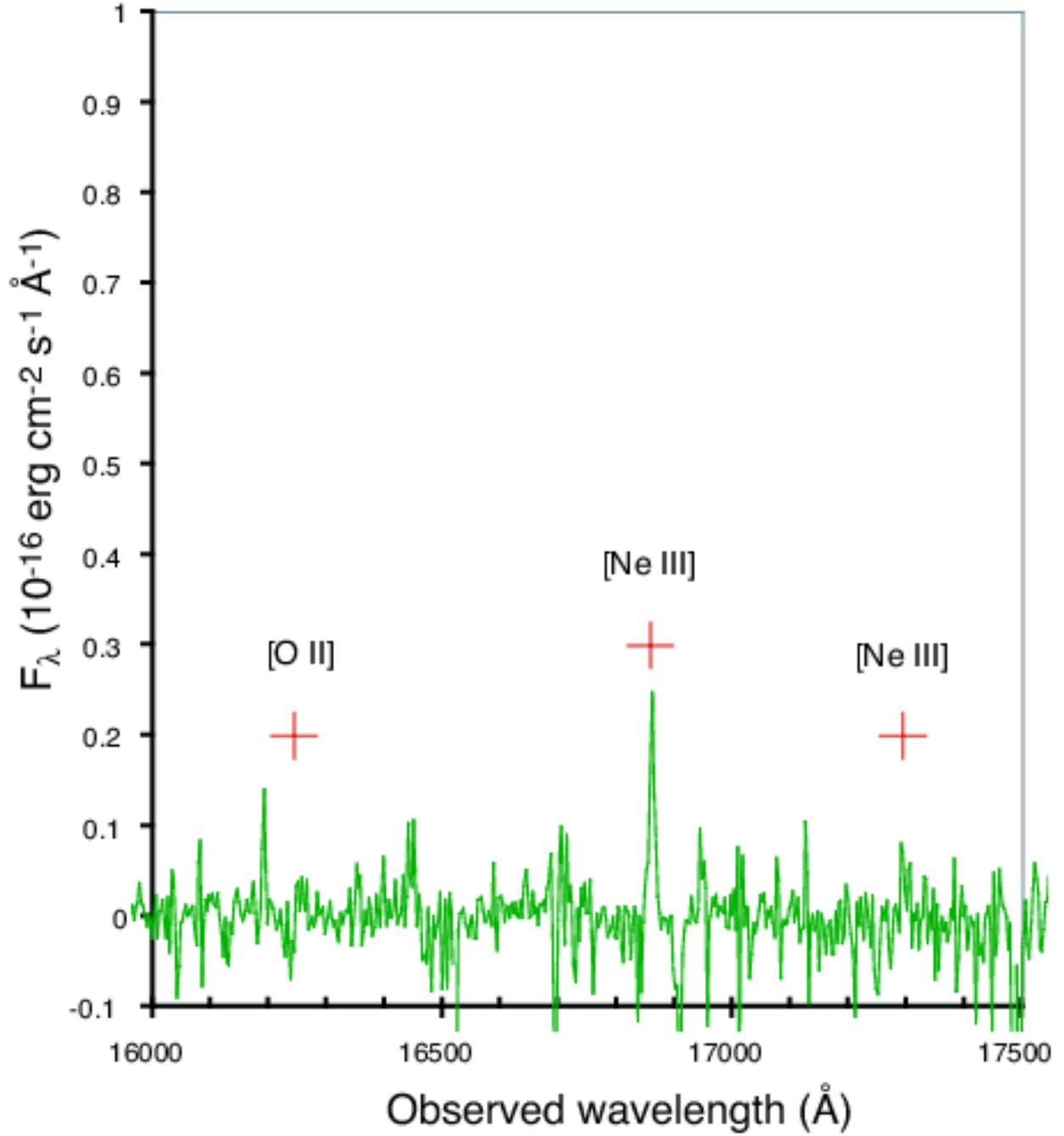}
\caption{NIRSPEC H--band spectrum of the arc. The ``+'' symbols mark the positions of the \oii\ doublet and the two \neiii\  lines.\label{nirspec-h}}
\end{figure}

\begin{figure}
\plotone{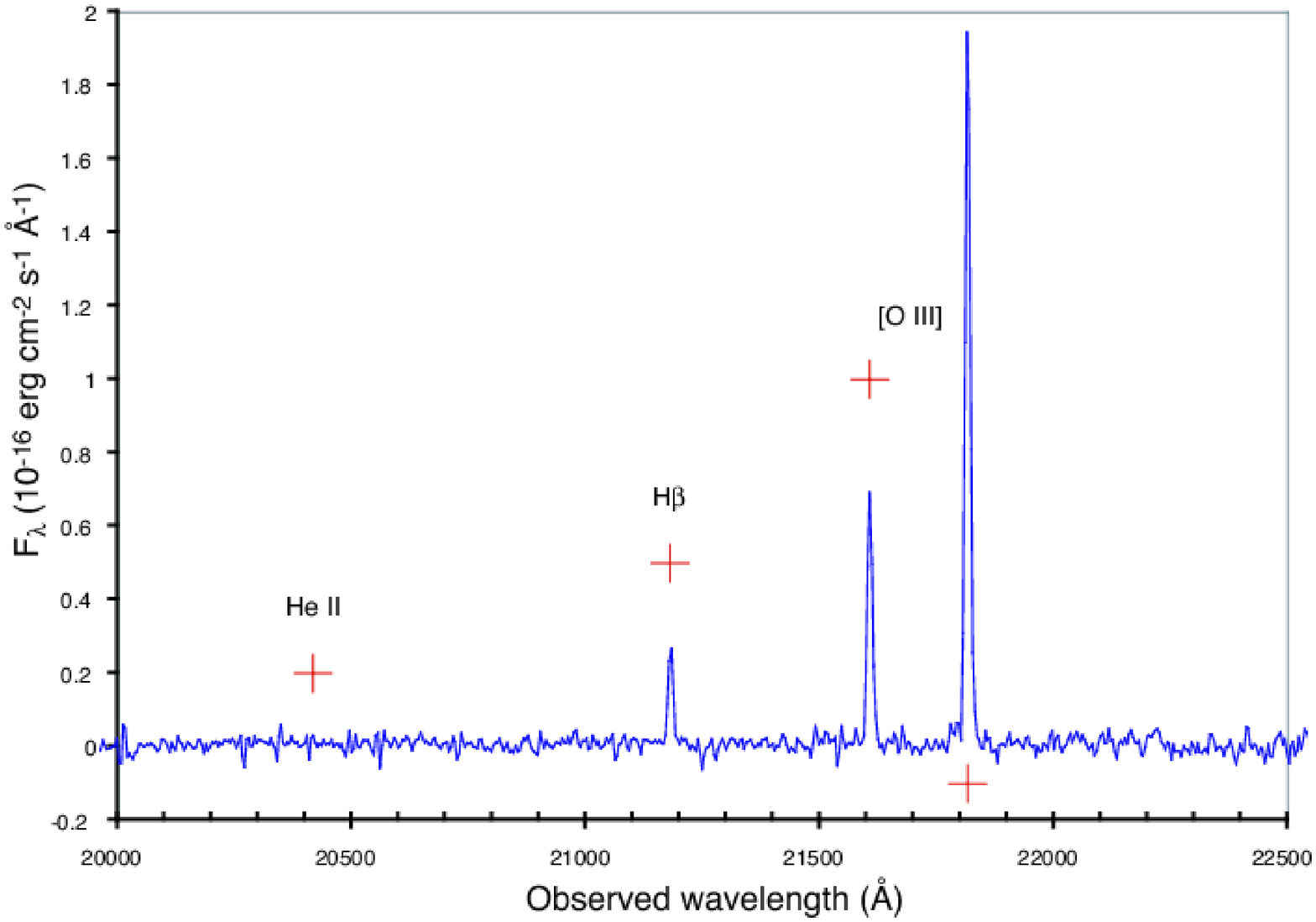}
\caption{NIRSPEC K-band spectrum of the arc. The ``+'' symbols mark the positions of the \heii, \hb\ and the \neb\ lines.\label{nirspec-k}}
\end{figure}

For measurements of the optical restframe spectrum, H and K band spectra were taken with the NIRSPEC instrument \citep{mcc98, mcc00} during UK service time with the Keck~II telescope on the night of UT 2001 November 20. The slit was $42 \times 0.76$\  arcsec in PA $46.5\degr$, aligned to include both the A \& B components. The spectrograph was used in its grating mode, giving a resolving power $R\sim2000$. The exposures in both the H (Nirspec-5; 1.413--1.808$\mu$m) and K (Nirspec-7; 1.839--2.630$\mu$m) bands were 450s. The data were reduced and wavelength calibrated using IRAF and STARLINK programs, in the standard manner.  The main steps in the reduction were (i) debias, (ii) remove cosmic rays and bad pixels, (iii) flat-field correction, (iv) geometric correction, (v) wavelength calibration, (vi) sky-subtraction.  Flux calibration was performed using observations of the standard star HIP~45994 compared to the expected spectrum of an F5V star ($T=6550$K) with K${}=7.66$. An aperture correction factor, determined by modelling light-loss due to a $0\farcs76$\ slit centred on a $0\farcs7$\ Gaussian seeing disc, was applied to the spectra. The H and K band regions containing the detected emission lines of \neiii, \hb\ and \neb$\lambda\lambda$4959, 5007 are shown in figures~\ref{nirspec-h} and \ref{nirspec-k}.

\clearpage

\clearpage

\placefigure{nirspec-h}
\placefigure{nirspec-k}

\subsection{Photometry}

Although the spectra were all taken in photometric conditions, the use
of different slit widths and orientations means that photometric
scaling factors must be applied in order to derive a consistent
calibration. This was accomplished using the following procedure in
order to derive an estimate of flux from the {\it sum} of the two
components A+B whose flux ratio, from the WFPC~2 data, is measured to
be A/B${}=1.67$. It is assumed \citep{hol01} that the spectra of A and
B are congruent.

\begin{itemize}
\item{The LRIS$_{l\rm ow}$\ spectrum of component A was scaled to represent A+B and then corrected for slit losses by a factor of 1.8.}
\item{ The LRIS$_{\rm med}$\ and the ESI spectra were then scaled such
    that the sum of the strong emission lines of \niv, \civ\ and
    \oiii\ matched the same sum in LRIS$_{\rm low}$.}
\item{The emission line fluxes from the three spectra were averaged, using equal weights, to produce the mean optical fluxes in table~\ref{line-fluxes}.}
\item{The NIR spectra were only corrected for the estimated slit loss.}
\end{itemize}

\begin{figure}
\plotone{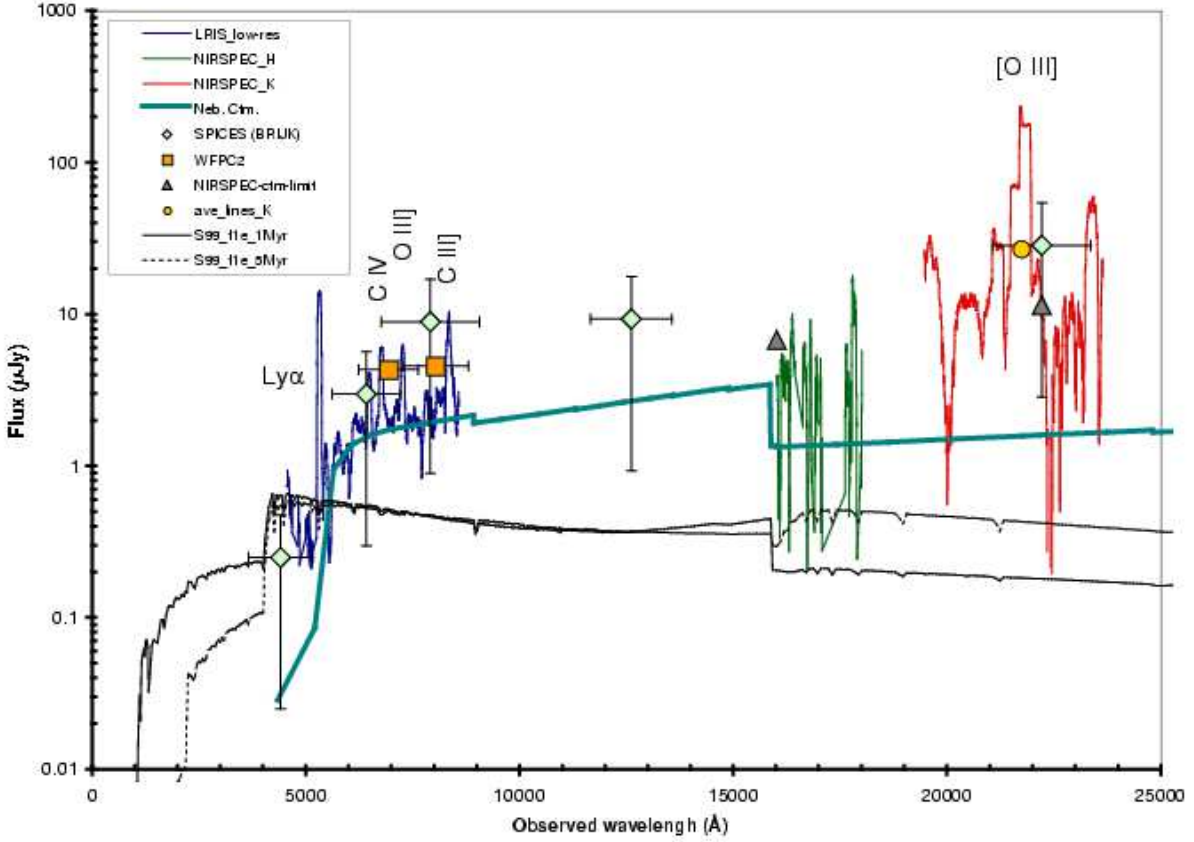}
\caption{The optical/NIR SED of the Lynx arc components A+B. Smoothed
  versions of the NIRSPEC and LRIS$_{\rm low}$\  data are plotted, scaled by the photometric correction factors discussed in the text. The HST aperture photometry (squares) is from the F814W and the F702W WFPC~2 images. 3$\sigma$\ upper limits to the H and K band continuum flux are plotted as triangles. The circle represents the \neb\ and the \hb\  fluxes summed within the K band. The diamonds represent preliminary imaging photometry in the B, V, I and K bands from the SPICES survey \citep{ste02, eisip}. The thick solid line is a model for the nebular continuum, scaled to the observed \hb\ flux from a gas at 20,000K assuming zero reddening. As a fiducial, we have plotted two synthetic SED from `STARBURST~99' \citep{lei99}. These represent instantaneous burst models of $10^7 M_{\odot}$\ at ages of one and five million years, scaled to the observed frame for the redshift of the arc and including a magnification factor of ten. See section \ref{continuum} for the other burst model parameters.\label{photometry}}

\end{figure}

The overall photometric data are summarised in
figure~\ref{photometry}, where we plot both the imaging and the
spectroscopy. Preliminary imaging photometry from the SPICES survey
\citep{ste02, eisip} is included but this could be contaminated by the
two nearby galaxies seen in figure~\ref{hst-image}. The HST imaging
photometry derived, using SExtractor {\tt magbest} \citep{ber96},  is given in table~\ref{hst-mags}.

\placefigure{photometry}

\placetable{hst-mags}

\clearpage

\begin{deluxetable}{lcc}
  \tablecaption{HST magnitudes (Vega) for components A and B of the
    arc. Component A in the F814W image may be contaminated by a
    residual cosmic ray event. Consequently, the value of the A/B flux
    ratio used in this paper is derived from the F702W data only. The
    magnitude uncertainties, based on the SExtractor {\tt MAG\_ISOCOR}
    results, are estimated to be $\sim0.05$.\label{hst-mags}}
  \tablewidth{0pt} \tablehead{ \colhead{Filter} & \colhead{A} &
    \colhead{B}} \startdata
  F702W & 23.32 & 23.88 \\

  F814W & 22.89: & 23.63 \\
  \enddata
\end{deluxetable}

\clearpage


\clearpage

\section{RESULTS}

The arc in the southern cluster \objectS\ is found at about $6''$ from the
brightest galaxy.  The dominant cluster to the north, \object,
has a much larger velocity dispersion --- $670 \pm 50$~\kms\ 
versus $430 \pm 20$~\kms --- \citep{hol01} and shows
several arclets but no clear multiple image.  Given the low velocity
dispersion of the southern cluster, it is reasonable to assume that
the northern one contributes significantly to the lensing of the $z =
3.357$ object.

The lensed system is composed of two clumps (A and B in figure~\ref{hst-image})
joined by a very faint, thin arc which can be followed for more
than $4''$ on either side. This forms an angle of about $30^\circ$ to the
tangent to the main cluster galaxy.  Moreover, two extremely faint
sub-clumps (C and D), possibly associated with the lensed galaxy, are
observed on the arc, symmetrically disposed with respect to images A
and B.  The overall system shows a clear mirror symmetry, suggesting that the source lies close to a fold caustic.  The two images
A and B do not have an obvious counter-image, which argues that they are highly magnified.  Indeed, the third unobserved image, in our
situation, should be of `Type~I' \citep{sch92}, and thus have a magnification (slightly) larger than unity.

In the restframe, our spectra extend from 1000\AA\ to 5400\AA\  with gaps from 2100--3500\AA\ and from 4100--4500\AA. There is a weak but clearly detected continuum in the ultraviolet, rising from $\la0.5$\ujy\ around \lya\ to $\sim2$\ujy\ near \ciii. This continuum is not, however, detected in the optical band from our NIR spectra: 3$\sigma$\ continuum limits from NIRSPEC are 5\ujy\ in H and 12\ujy\ in the K band.

The emission spectrum is unusual in terms of the relative line intensities. In particular, the intercombination lines of \niv, \ciii\ and \oiii\ are very strong while forbidden \oii\ is undetected and much weaker than \neiii. The absence of \nv\ and the weakness of \niii\ suggests that the source of ionization is blackbody-like rather than a power-law extending to high energies. \heii\ $\lambda$1640 is detected but we do not see \heii\ $\lambda$4686 in our NIR data. The possibility that such an emission spectrum could arise by photoionization from a `filtered' AGN continuum in the outer parts of an obscured quasar host galaxy is discussed by \citet{bin03}. They consider mimicing a hot black-body by partially absorbing a power-law spectrum with high ionization parameter gas close to an AGN. Such an explanation is not pursued further in this paper.

A spectrum of this nature, and we have been able to find nothing that
closely resembles it in the available UV spectral archives or the
literature, suggests that we are dealing with a compact (high
ionization parameter $U$) low density gaseous nebula, ionized by a hot
($T_{\rm eff} \sim 8\times10^4$K) thermal source. The lines are narrow (FWHM${}\la{}$100~\kms) although \lya\ is broader and self-absorbed. \civ\ has a complex profile that also suggests absorption of the emission doublet. Our optical spectra show that the \niv, \civ\ and \heii\ lines have broader, low intensity wings while the \oiii\ exhibits a narrow base. The \hb\ and \neb\ are unresolved in our spectra but they show no signs of a broad base.

\subsection{Continuum}
\label{continuum}

The restframe UV continuum rises longward of \lya\ and there is no evidence for an underlying blue stellar population unless it suffers significant reddening. Since the emission line equivalent widths are so large, we have calculated the expected nebular continuum, predominantly  the hydrogen 2-photon emission in this spectral range, from the observed strength of \hb. The coefficients are from \citet{all84}, assuming $T_e = 20,000$K, a temperature justified by the photoionization modelling presented in section~\ref{photoionization}. The calculated continuum is plotted in figure~\ref{photometry} and it is a reasonable fit to both the shape and level of the observed SED in the restframe UV, suggesting that the nebula is essentially reddening-free.

To gain some quantitative insight into the limits that can be placed on the nature of any underlying stellar population, we have plotted two fiducial synthetic SED on figure~\ref{photometry}. These are from `STARBURST 99' \citep{lei99} and represent instantaneous burst models of $10^7 M_{\odot}$\ with a Salpeter IMF with $\alpha = 2.35$, an upper and lower mass cut-off of $100M_{\odot}$\ and $1M_{\odot}$\ respectively and a metallicity of $Z/Z_{\odot} = 0.05$. They include both the stellar and nebular continuua and their ages are one and five million years. The model luminosities have been scaled to observed fluxes from the redshift of the arc, using the cosmological model defined in section \ref{ionizing}, and we have multiplied by a lens magnification factor of ten.

\subsection{Line spectrum}
\label{line-spectrum}

Emission lines have been measured from the different spectra and averaged, using the procedure described above, to represent the sum of fluxes from components A and B. The results are presented in table~\ref{line-fluxes} together with the velocity dispersions of the narrowest components fitted to the ESI data for the stronger lines.

\placetable{line-fluxes}

\begin{figure}
\plotone{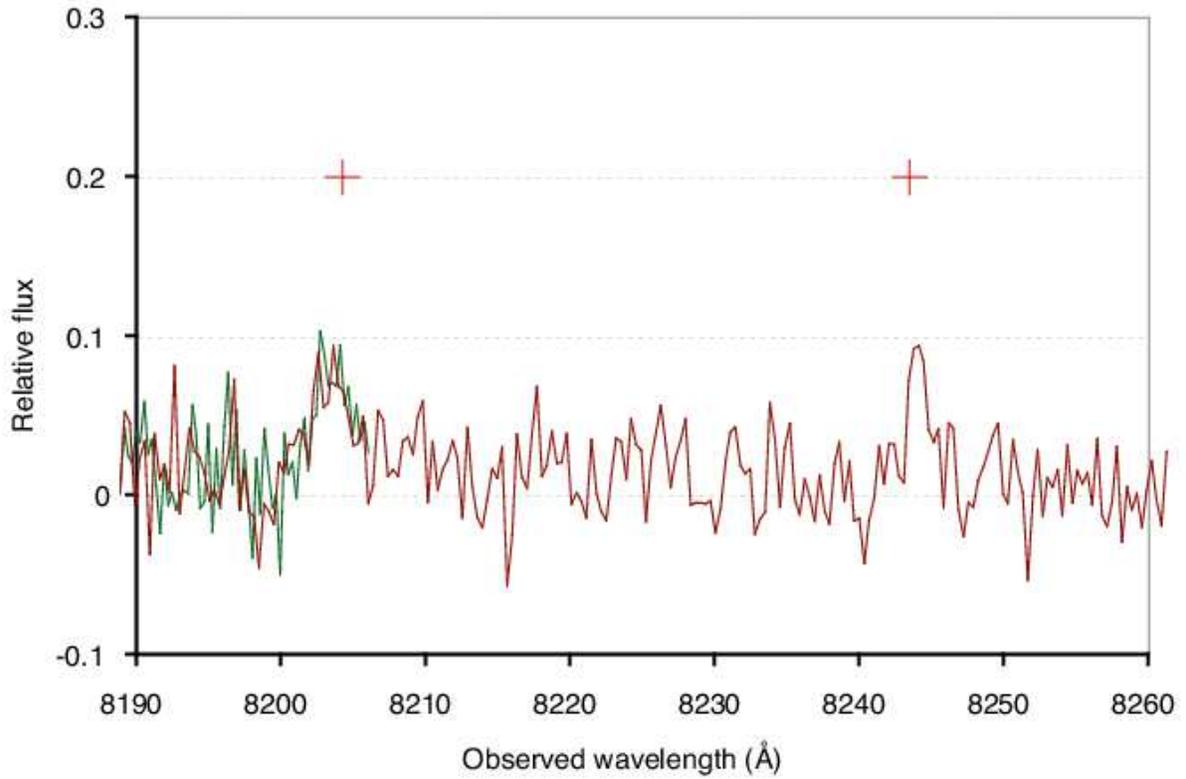}
\caption{The \siiii\ $\lambda\lambda$ 1883, 1892 region in the ESI spectrum. The ``+'' symbols mark the redshifted position of the doublet. The shorter wavelength line is seen in two spectral orders.\label{silicon3}}
\end{figure}

The \siiii\ doublet $\lambda\lambda$ 1883, 1892 seen in the ESI spectrum merits some comment.  It is underpredicted by all of our photoionization models using scaled solar abundances (see below) but the 1883\AA\ line is seen in two ESI orders and, while close to the formal 3$\sigma$\ detection limit, the 1892\AA\ feature is at precisely the correct wavelength to be the second component of the doublet (see figure~\ref{silicon3}). We, therefore ascribe a high confidence level to the detection.

\placefigure{silicon3}

\clearpage

\clearpage

\begin{figure}
\plotone{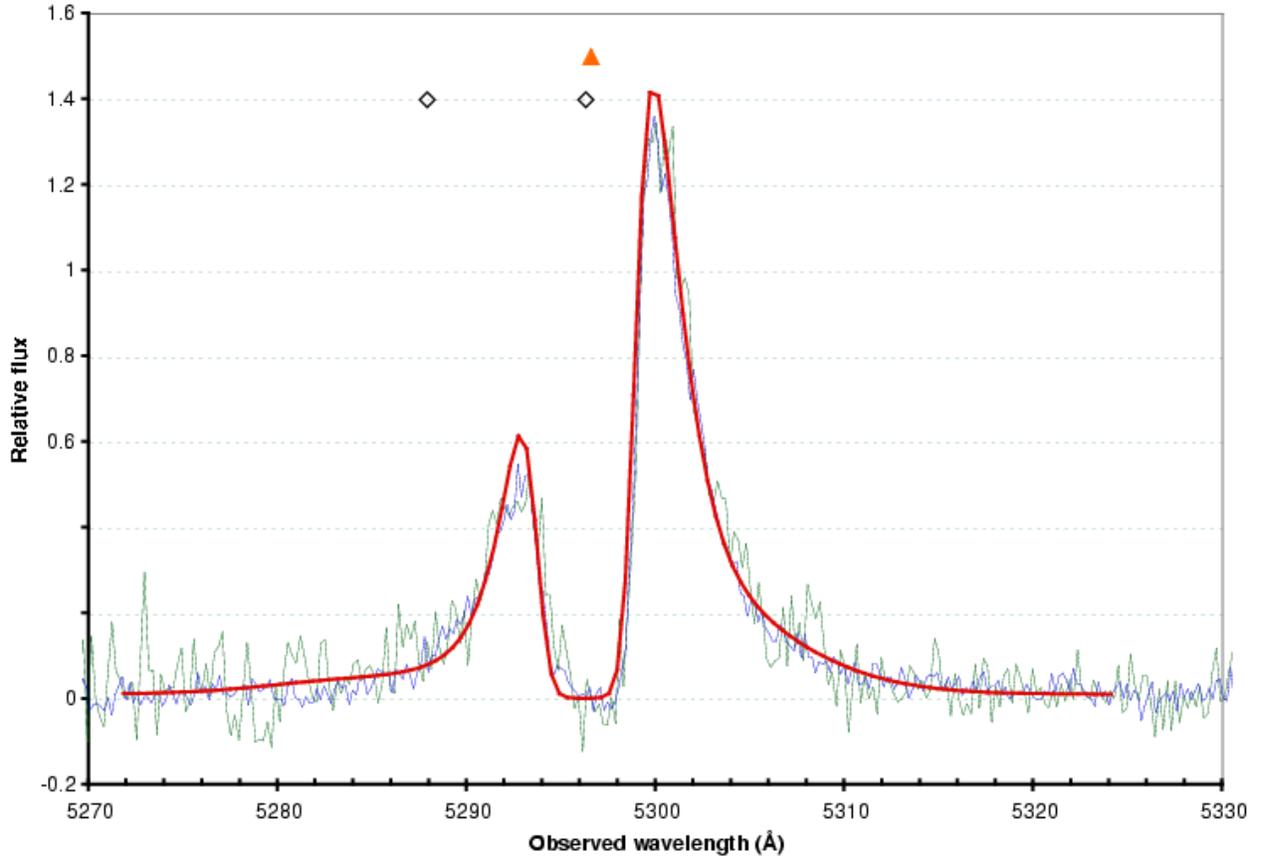}
\caption{Two orders from the ESI spectrum of \lya. The thick line shows a fit using two emission and two absorption components modelled as pure Gaussians. The fit parameters are given in table~\ref{abs-fit}. The velocity centres of the emission (filled symbol) and the two absorption components from the model are marked.\label{lya-fit}}
\end{figure}

\begin{figure}
\plotone{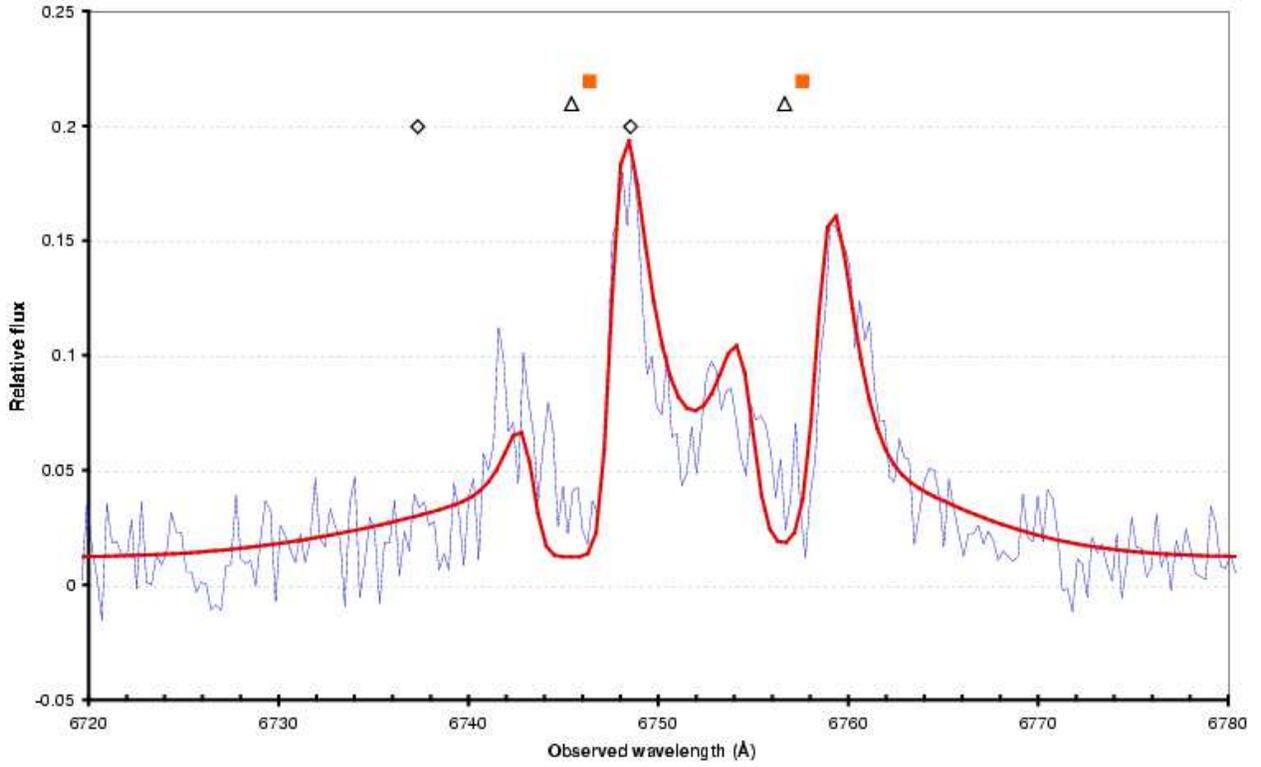}
\caption{The ESI spectrum of \civ. The thick line shows a fit using two emission and two absorption doublet components modelled as pure Gaussians. The fit parameters are given in table~\ref{abs-fit}. The velocity centres of the emission (filled symbols) and the two absorption components from the model of the doublet are marked.
\label{civ-fit}}
\end{figure}

Both \lya\ and \civ\ are affected by absorption which we have modelled
using pure Gaussian line emission and absorption coefficients. When
fitting the ESI data, the model components are broadened in quadrature
using an instrumental width derived from sky and arc lines and taking
the form $\sigma_{\rm inst} = -0.003\lambda$(\AA)${} + 51$~\kms. The fit to \lya\ is shown in figure~\ref{lya-fit} and the parameters for the two emission and two absorption components are given in table~\ref{abs-fit}. These four components are the minimum number required for a satisfactory fit and show the presence of an outflowing wind with a low velocity component at  $-$15~\kms\ having an \hi\ column density of $\sim10^{15}$~\cmmt. The \civ\ doublet is similarly modelled with each component having two emission and two absorption components (figure~\ref{civ-fit}) with the doublet separation and oscillator strength ratio constrained to their theoretical values. These fit parameters also appear in table~\ref{abs-fit} together with the H$^0$\ and C$^{3+}$\ column densities derived from $\tau_0$\ and $\sigma$. The \civ\ fit is less well constrained due to the lower s/n of the data for this weaker line. The value of $\tau_{0, 1548}$ is, however, required to be $\geq 4$\ in order to match the steepness of the red edges of the absorption troughs. Note that for both \lya\ and for \civ, the two emission components have no relative velocity offset. The velocity offsets of the absorption components in the two species are not identical. Such an effect could result from an ionization gradient in the flow. The absorption correction factors for \lya\ and \civ\ used in table~\ref{line-fluxes} are derived by setting the absorption optical depths equal to zero in these fits.

\placefigure{lya-fit}

\clearpage

\clearpage

\placefigure{civ-fit}

\clearpage

\clearpage

\placetable{abs-fit}

\clearpage

\begin{deluxetable}{lccccc}
\tablecaption{The emission and absorption components in \lya\ and \civ. The emission profiles are pure Gaussians as are the line absorption coefficients. The velocity offsets with respect to the redshifted line centres, the instrument-corrected Gaussian velocity dispersions, the emission intensities (normalised to a total of unity for the summed components), the line centre absorption optical depths and the absorbing column densities are given.\label{abs-fit}}
\tablewidth{0pt}
\tablehead{
\colhead{Component} & \colhead{$\Delta v$}   & \colhead{$\sigma_{\rm corr}$}   &
\colhead{Intensity} & \colhead{$\tau_0$}& N\\
& \colhead{\kms} & \colhead{\kms} & normalised&  &\colhead{10$^{15}$~cm$^{-2}$}}
\startdata
 \lya & & & & & \\
e$_1$ & 0 & 150 & 0.71 &  & \\
e$_2$ & 0 & 400 & 0.29 & & \\
a$_1$& $-$15 & 63 & & 9 & 1.05\\
a$_2$ & $-$490 & 270 & & 1.2 & 0.60\\
\tableline
& & & & & \\
\civ& & & & & \\
 e$_1$ & 0 & 80 & 0.43 & & \\
e$_2$ & 0 & 350 & 0.24 & & \\
 a$_1$& $-$40 & 35 & & 7.5 & 0.83\\
 a$_2$ & $-$400 & 400 & & 0.8 & 1.02\\
\enddata
\tablecomments{For \civ, the emission doublet is assumed to have an optically thin 2:1 ($\lambda$ 1548:1551) intensity ratio: the fitted parameters and normalised intensities refer to the stronger line (1548\AA). For the absorption, it is the line absorption coefficients that have the 2:1 ratio.}
\end{deluxetable}

\clearpage

\section{DISCUSSION}

In section~\ref{lensing}, the lens modelling is discussed with the principal aim of computing the magnification that will allow the estimation of the true observed brightness of the source.  Using a grid of photoionization models in conjunction with the observed emission line ratios, we discuss in section~\ref{photoionization}\ the nature of the source: physical conditions, nebular element abundances and the characteristics of the ionizing source. The absorption components seen in \lya\ and in \civ\ allow us to deduce some properties of the wind for comparison with low-redshift \hii\ galaxies. Finally, in section~\ref{ionizing}\ we compare the observed line luminosity, using a current `standard' cosmology, with the value prediced from the $L_{H\beta} - \sigma_{H\beta}$\ calibration of local \hii\ regions/galaxies from \citet{mel99}.

\subsection{Lensing analysis}
\label{lensing}

The primary goal of the lensing analysis in this study is to constrain
the magnification of the lens.  The modeling has been carried out
using two guiding principles. Firstly, we note that the system is
intrinsically complex because of the cluster environment.  Hence, an
accurate description of the mass distribution should include several
components, corresponding to the main galaxies that are contributing
to the lensing signal. Secondly, the number of constraints available
from the observed images is rather small.  In particular, using the
observed positions, ellipticities and fluxes of the two main images (A
\& B) we have 5~free parameters for the lensing model.\footnote{In
  fact, we have 5 parameters for each image (2 from the position, 1
  from the luminosity, and 2 from the ellipticity of the image), but
  we have also to use 5 parameters to describe the source.  Note that,
  because of the conservation of the surface brightness in lensing,
  using the size of the source would not gain us any additional
  information.}

These two competing facts put strong constraints on the families of
models we can use.  With only two images, we cannot use more than 5
parameters; on the other hand, since the system is complex, we need to
take into account the effects of the several masses that might
contribute to the lensing.  A standard method, which we adopt here, is
to model the \textit{collective\/} effect of nearby galaxies as an
external shear (see, e.g., \citealp{kee97}).  More formally, it can be
shown that, within a given region of sky, the gravitational lensing
effect of external masses can be described using a multipole expansion
(see Schneider \& Bartelmann 1997). In this framework, the external
shear represents a quadrupole term, this being the first observable
multipole.  We also stress that in many strong lensing studies the
external shear is required for a good fit seems to be larger than one
would expect by considering the influence of the nearby masses (see,
e.g., \citealp{wit97}).

\begin{figure}
  \plotone{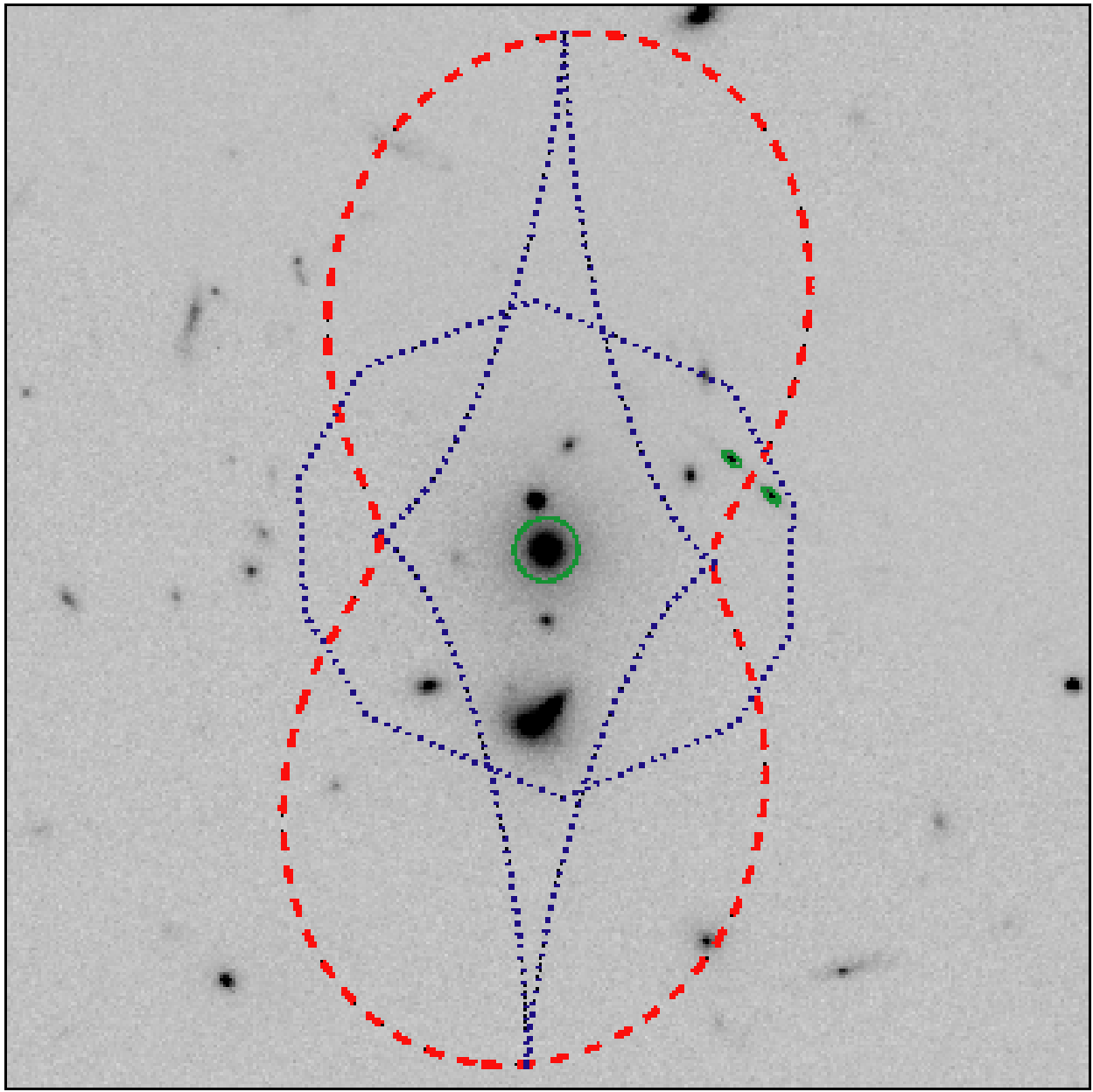}
  \caption{The lens model 1 (table~\ref{lens-table}).  The dashed
    curve represents the critical line on the lens plane and the
    dotted curves the caustics on the source plane.  The main lensing
    galaxy and the two images A and B are marked with continuous
    lines. This figure has the same size and scale as
    figure~\ref{hst-image} but a different centre.\label{lens-model}}
\end{figure}

A simple quantitative analysis can be used to better understand the
lens.  Assuming that the system is symmetric and that no external
field is acting upon it, we can obtain a first estimate of the mass
inside the Einstein radius of the system.  The definition of the
Einstein radius is in our case rather arbitrary since, as mentioned
above, the apparent centre of the arc is far away from the centre of
the brightest galaxy in \objectS\ (the centre of
figure~\ref{lens-model}).  In any case, assuming an Einstein radius of
about $6''$ (i.e., the observed angular distance between the arc and
the brightest cluster galaxy), we obtain a projected velocity
dispersion for the lens of about 550~\kms.  This is likely to be too
high for a single galaxy and too small when compared with the velocity
dispersion of the cluster.  Hence, this suggests that the lens system
can be explained only if we take into account both the strong
gravitational field of the central galaxy and the external tidal field
of the cluster.  Notice that the detailed lens modeling presented
below is independent of the simple analysis discussed here.

\citet{mao98} have discussed in detail the impact of substructures on
the lens models.  They note how, in several cases, simple lensing
models are able to account for the observed configurations but fail to
reproduce the observed fluxes ratios between the various images.  The
conclusion is that small-scale substructure could play a significant
role.  The effect of substructures is particularly important in
highly-magnified images where even a small change in the local density
(for example, due to a compact object such as a globular cluster) can
have large effects on the observed fluxes; more recent studies reach
the same conclusion (see, e.g., \citealp{bra02} and references
therein).

We have modeled the lens as a single elliptical profile, centred on
the dominant cluster galaxy (figure~\ref{lens-model}). Contributions
from other galaxies have been represented by an external shear.  The
mass profile of the lens was taken to be a pseudo-isothermal profile
(see, e.g., \citealp{kee98} with velocity dispersion $\sigma_0$, core
radius $r_0$, and truncation radius $r_1$.  More specifically, the
surface mass density of the lens (i.e., the mass density projected
along the line of sight) is taken to be

\begin{equation}
  \Sigma(x, y) = \frac{\sigma_0^2}{2 G} \frac{r_1}{r_1 - r_0} \left(
  \frac{1}{\sqrt{r_0^2 + \rho^2}} - \frac{1}{\sqrt{r_1^2 + \rho^2}}
  \right) \; ,
\end{equation}

\noindent where $\rho$ is the elliptical coordinate for an ellipse of axis $(1
+ e, 1 - e)$: 

\begin{equation}
  \rho^2 = \frac{x^2}{(1 + e)^2} + \frac{y^2}{(1 - e)^2} \; .
\end{equation}

\noindent Note that we used a model with an elliptical mass distribution rather
than an unphysical elliptical potential (see \citealp{kas93} for a
discussion on this point).  This mass distribution is characterized by
a total mass (in the limit of small ellipticities, $e \to 0$)

\begin{equation}
  M_\mathrm{tot} = \frac{\pi}{G} \sigma_0^2 r^2_1 \; .
\end{equation}

\noindent The lens parameters were obtained by fitting the predicted image
positions, fluxes and ellipticities to the observed data.  The fitting
procedure was carried out first on the source plane \citep{kay90}.
Given a lens model, the observed positions of each image were
projected onto the source plane and compared to each other.  In our
analysis, we extended this technique to include the observed fluxes
and ellipticities.  The final fit was then carried out by inverting
the lens equation and fitting directly the observed quantities on the
lens plane. The lens inversion was performed using a technique similar
to the one described by Keeton (2001).  The final minimization was
done using a simulated annealing algorithm (\citealp{ann83};
\citealp{ann84}).  This method, which is based on an analogy with a
thermodynamic process (the formation of crystals as a liquid
\textit{slowly\/} freeze), has proved to be very effective in dealing
with minimization of functions in high-dimension spaces, when the
presence of local minima can create severe problems to standards
algorithms (see, e.g., \citealp{NR}).  In the following we briefly
present the final results of the lensing analysis.

\placetable{lens-table}

\begin{table}
\centering
  \begin{tabular}[c]{lcccc}
    \hline\hline
    Parameter & Model 1 & Model 2 & Model 3 & Model 4\\
    \hline
    Lens ellipticity $e$ & $0.06$ & $0.02$ & $0.66$ & $0.09$ \\
    Position angle $\theta$ [degrees] & $74$ & $45$ & $34$ & $48$ \\
    Vel. disp.\ $\sigma_0$ [$\mbox{km s}^{-1}$] & $643$ & $596$ & $427$ &
    $483$ \\
    External shear $\gamma_1$ & $0.52$ & $0.44$ & $0.35$ & $0.37$ \\
    External shear $\gamma_2$ & $0.01$ & $-0.01$ & $-0.47$ & $-0.26$ \\
    \hline
    Total $\chi^2$ & $5.33$ & $26.87$ & $9.54$ & $27.61$ \\
    Degrees of freedom & 2 & 4 & 6 & 7 \\
    Magnifications $\mu_{A,B}$ & $(6.57, 9.85)$ & $(6.17, 9.24)$ &
    $(4.77, 7.14)$ & $(3.27, 4.89)$ \\
    \hline\hline
  \end{tabular}
  \caption{The relevant parameters for the four lens models described
    in the text.}
  \label{lens-table}
\end{table}

\placefigure{lens-model}

For each lens model the fitting algorithm was run several times in
order to check the reliability of the results obtained.  Indeed, the
minimization procedure, based on a simulated annealing method, has an
external parameter, the `temperature' of the system, that needs to be
carefully tuned (see \citealp{NR} for a brief discussion on the
simulated annealing algorithm). For example, a low temperature can
prevent the algorithm from finding the global minimum of the $\chi^2$
function.

In order to better understand the degeneracies of the system, we tried
several models. In almost all cases, we were able to reproduce the
observed positions, fluxes and ellipticities of the images with good
accuracy (see table~\ref{lens-table}).  Clearly, we do not claim that
all the models are an accurate description of the system. Rather, we
provide a number of models that give good fits to the data in order to
show the possible range of magnifications for the system.  All models
require a very large external shear.  Interestingly, the external
shear of many models (see in particular models 1, 2, and 4 in
table~\ref{lens-table}) is approximately aligned with the direction of
the Northern cluster.  Assuming a velocity dispersion of $670 \pm
50$~\kms \citep{hol01}, we find that about one half of the external
shear required by our models can be simply explained by this cluster;
the remaining external shear is probably due to other galaxies in
\objectS.

The simplest model tried on the system is a singular isothermal
elliptical lens (SIE) centred on the main cluster galaxy in \objectS\ 
with an additional external shear (model 1 in table~\ref{lens-table})
This model is characterized by five free parameters: the lens velocity
dispersion $\sigma_0$, ellipticity $e$, position angle $\theta$\ and
the two components of the external shear, $\gamma_1$ and $\gamma_2$.
Although the model is very simple, the final fit is very good; the
only significant residuals come from the fit to the flux (which
contributes to the total $\chi^2$\ of $5.33$).  The best model has a
round lens and a rather large external shear. This produces a faint
third image (about 10 times fainter than image A) which would be
difficult to detect in our data.  Figure~\ref{lens-model} shows the
best fit configuration.

In order to reduce the external shear, we tried to add a significant
`penalty' on the $\chi^2$ for large external shears (model~2).  This,
however, did not reduce significantly the shear and produced larger
errors on the fluxes.  Interestingly, because of the well known
degeneracy between the external shear and the lens ellipticity (see,
e.g., \citealp{kee97}), the final fit obtained was qualitatively
different from model~1.

The inclusion of 4 more images (see figure~\ref{hst-image}) in the
fitting algorithm led to model~3.  These extra images include the
mirror-symmetric components C and D as well as fainter substructure
within A and B.  The best solution has a low $\chi^2$, a very faint
third image but a large lens ellipticity and external shear.

In model~4, we tried to use a candidate third image labelled T in
figure~\ref{hst-image}.  This object, from groundbased observations
\citep{ste02, eisip}, is found to have a color consistent with our
source and is in a good position to be a candidate third image.  In
all models obtained in this configuration, further images are either
not present or extremely faint.  The best fit model shows significant
scatter on the fluxes and on the ellipticities and thus has a large
$\chi^2$.

Finally, we also tried to relax the position of the lensing galaxy
around the observed position; this, however, never significantly
improved the fit.

In summary, we find a relatively wide range of models that
successfully describe the system.  The total magnification for the
images A$+$B in our models span the range $8.1$--$16.4$.  Hence,
although the present study is not able to completely constrain the
system, there are indications that the source could be strongly
magnified by the lens.  The class of acceptable models would be
narrowed should a third image be found. For the following discussion,
we should be aware that, by being so close to a caustic, some part of
the source may be more highly magnified than the average. This could
affect the interpretation of line ratios and widths in terms of
spatially-averaged models.

\subsection{Photoionization modelling}
\label{photoionization}

Photoionization modelling is carried out using the \map\ code
\citep{fer97}. For the exploration of the spectrum discussed in this
paper, we have assumed that the gas is ionized by a pure black body.
In effect, the stellar cluster is replaced by a single ionizing star
with a temperature $T^*$ \citep{eva85}, based on the notion that the
ionizing spectrum is not strongly dependent on the slope of the IMF.
This is a reasonable approximation in stellar clusters since the
number of ionizing photons increases so rapidly with the temperature
of the hottest stars \citep{sea71}. The acceptability of the black
body approximation as a representation of the real SED of a massive
star is, of course, strongly dependent on the stellar metallicity
although for metal-free stars \citep{bro01} they are very similar.

Deducing the true nature of the ionizing stellar cluster from the
nebular emission alone is difficult. There are, however, several
important clues that indicate the presence of massive, luminous and
probably very low metallicity stars. These are:

\begin{itemize}

\item The color temperature of the ionizing source is much higher than seen in \hii\ regions in the local universe. The ionizing flux in typical Galactic compact \hii\ regions are typically characterized by stellar model effective temperatures $T_{eff} \lesssim$ 40,000K \citep{lum03} which correspond to somewhat lower color temperatures.

\item The ionization parameter, $U$ is also higher than seen in local \hii\ regions.

\item The nebula emits \heii\ $\lambda$1640.

\item No stellar continuum is observed between \lya\ and the visible spectrum.

\item The cluster has a very high ionizing photon luminosity yet the nebular lines are narrow, indicating a low gravitating mass (see section~\ref{ionizing}).

\end{itemize}

While these observations do not uniquely diagnose the presence of Population~III stars, they are all qualitatively consistent with the predictions of the models of such stars and their surrounding nebul\ae. In particular, the question of the actual time-averaged ionizing properties of system containing massive Population III stars, including the effects of evolution away from the ZAMS and assumptions regarding the IMF are addressed by \citet{sch02}. 

The \ciii\  emission line ratio indicates that the ionized gas density is at or below the low density limit for this doublet ($\sim 1000$~cm$^{-3}$, \citealp{ost89}). In our models, we assume $n_e = 50$~cm$^{-3}$, ensuring that there are no significant collisional de-excitation effects. An increase in density would result merely in a decrease in the ionization parameter $U$. Our models are isobaric which means that we set the density at the illuminated face of the cloud and it increases with depth to keep the pressure constant.

We have assumed a plane-parallel geometry, modelling the emission from a single `cloud' ionized by a central cluster of stars. The output spectrum can be considered as the integrated emission from many clouds distributed spherically around the ionizing source. In dust-free models, different geometries will be represented by different effective values of the ionization parameter $U$. In the presence of dust, however, the geometry will determine the photon escape probabilities and so will have a much stronger effect on line ratios, especially those involving \lya\ and \civ. Since models of low redshift \hii\ regions generally employ a spherical geometry, it should be noted that our values of $U$ are not directly comparable.

Our photoionization models explore a parameter space consisting of the
ionizing source black body temperature $T_{\rm BB}$, the ionization parameter $U$ and the gas metallicity $Z$. The metallicity has been scaled in proportion to the relative solar abundances \citep{and89} and assumes no depletion onto dust. Such a representation of the element abundance distribution is probably quite inappropriate for a source at this epoch but we do not explore this part of the parameter space in this paper. Our objective is to find a `fiducial' model that represents the general nature of the emission line spectrum and to discuss how departures from it might be addressed. 

The color temperature of the photoionizing source is relatively
well-determined by the nitrogen spectrum. The dominance of \niv\ over
both \nv\ and \niii\ restricts $T_{\rm BB}$ to $80,000\pm10,000$~K. Such an ionizing SED would dominate the continuum spectrum shortward of \lya\ but, in our observed spectral range, would be much weaker than the nebular continuum. This is consistent with the result presented in figure~\ref{photometry} --- see also: \citet{pan02}.

There are two line ratios that are sensitive to metallicity but
essentially independent of stellar temperature and only weakly
dependent on $U$: \heii/\civ\ and \neb/\hb. The observed ratios both
exclude very low metallicities ($\lesssim 0.01 Z_{\odot}$) which would result in
very strong H and He lines relative to \civ\ and \neb. Similarly, $Z  \gtrsim  0.05 Z_{\odot}$
 is excluded by the observed value of \neb/\hb\ 
in this regime where oxygen is not the dominant coolant and the ratio
depends directly on metallicity. Consequently, we assume a gas-phase
metallicity of $Z  = 0.05 Z_{\odot}$\ with an uncertainty of about a factor of two.
The
line spectrum in general is best reproduced with a high ionization
parameter, $\log{U} = -1.0\pm0.5$.  Such a high value of $U$
distinguishes the spectrum from that of a planetary nebula which can
have a similar stellar color temperature. The line spectrum produced
by this fiducial model ($T_{\rm BB}=80,000$K, $U=0.1$, $Z/Z_{\odot}=0.05$) is
presented in table~\ref{line-fluxes}.

Photoionization models with these parameters always predict weak
\siiii\ lines: our fiducial model implies a Si/C ratio approximately forty times larger than the solar ratio. This problem can be alleviated by decreasing the
ionization parameter but at the cost of essentially eliminating the \niv, \oiii\
and \civ\ lines. Similarly, decreasing $T_{\rm BB}$\ will strengthen \siiii\ but weaken the high ionization lines. 
The minimum Si/C ratio we can obtain by decreasing $T_{\rm BB}$\ to 60,000K is around five in solar units. This quantitative conclusion should be treated with some caution due to the simplicity of our photoionization model, e.g., the use of a black body SED in place of a real stellar continuum.

\subsection{The ionizing source}
\label{ionizing}

The rate of ionizing photon production needed to power the arc can be
estimated from the de-magnified \hb\ flux. Using a cosmology with $H_0
= 65$~km s${}^{-1}$ Mpc${}^{-1}$; $\Omega_\mathrm{M} = 0.3$\ and $\Omega_\Lambda =
0.7$\ with the source redshift of $z = 3.357$, we compute a luminosity
distance of $D_\mathrm{l} = 3.13\times 10^4$ Mpc. The (hydrogen) ionizing
photon luminosity $Q_\mathrm{ion}$\ is then given by:

\begin{equation}
Q_\mathrm{ion} \simeq \frac{ 4 \pi D^{2}_\mathrm{l} F_{H\beta}  } { \mu h \nu_{H\beta}}  
\frac{\alpha_\mathrm{B}} {\alpha^\mathrm{eff}_{H\beta}}  \; ,
\end{equation}

\noindent where $\mu$\ is the lens magnification and the $\alpha_\mathrm{B}$\ and $\alpha^\mathrm{eff}_{H\beta}$\ are respectively the total and effective \hb\ case B hydrogen recombination coefficients for $T_e = 20,000$K \citep{ost89}. Using $\mu_{A,B} = 10$\ and $F_{H\beta}$\ from table~\ref{line-fluxes}, we get $Q_\mathrm{ion} \simeq 1.6 \times 10^{55}$~ph s$^{-1}$, assuming a nebular covering factor of unity.

This number can be used to constrain the nature of the ionizing stellar population. The number of massive stars needed to provide this photon flux depends in detail on the cluster IMF and on the stellar metallicity. The stellar color temperature we deduce from the emission line spectrum is very high and argues for a low stellar metallicity, possibly much lower than the nebular metallicity which may be affected by stellar winds and earlier supernova explosions. Using the Population~III stellar models given by \citet{sch02}, the appropriate range for $\log{Q_\mathrm{ion}}$ lies between about 49 and 50, indicating that we are dealing with up to 10$^6$\ massive stars.

The observed SED can also place constraints on the stellar population subject to uncertainties in the extinction to the stellar cluster. Figure~\ref{photometry}\ includes STARBURST~99 models (see section~\ref{continuum}) that illustrate the level and shape of the SED expected for a very young cluster of $10^7 M_{\odot}$\ with a normal Salpeter IMF lensed by the cluster of galaxies. An unreddened cluster of this mass would be marginally detected in our optical data but would only produce a $Q_\mathrm{ion} \simeq 8\times10^{53}$ ph s$^{-1}$, a factor of twenty below the estimated requirement.

The modest value of the velocity dispersion measured for the \oiii\ and the \ciii\ lines of $\sigma \simeq 33$~\kms\ places an effective limit on the gravitating mass within the line emitting region. If we use the calibration of the $L_{H\beta} - \sigma$\ relation for \hii\ galaxies by \citet{mel99}

\begin{equation}
\log L_{H\beta} = 5 \log \sigma - \log (O/H) + 29.5  \; ,
\end{equation}

\noindent we predict the intrinsic \hb\ luminosity to be $\log L_{H\beta}  = 41.5$\ for the nebular oxygen abundance of $12 + \log (O/H) = 7.6$\ that is implied by our photoionization models. This is about 150 times smaller than the observed value of 43.63 and is inconsistent with our estimate of the lensing magnification of only $\mu_{A,B} = 10$. 

There are many uncertainties in this argument, not least the extreme extrapolation of the $L_{H\beta} - \sigma$\ relation to low metallicity. A crucial issue may be the nature of the metallicty term in this relation whose value was determined empirically by \citet{mel88} for local \hii\ regions and galaxies. The physics of the metallicity dependence is to do with the luminosities of the ionizing stars ($Q_\mathrm{ion}$) rather than the chemical composition of the surrounding nebula. It is the metallicity of the stars that determines their IMF and mass-to-light ratio and, at early epochs as we have noted above, this may be considerably lower than the wind and supernova-polluted surroundings. Whatever the solution of this discrepancy turns out to be, there is a strong indication that the Lynx arc has a very low mass-to-light ratio implying an IMF dominated by massive stars.

Emission line velocity widths and characteristic sizes for faint sources with $z \sim$ 0.19 to 0.35 were measured by \citet{guz96} who found Gaussian velocity dispersions of $\sigma \sim$ 30 to 50 \kms\ and inferred masses of $\sim 10^9 M_{\odot}$. For a higher redshift sample of luminous, compact, blue galaxies with $z \sim$ 0.4 to 1.2, \citet{guz03} used multi-broad-band photometry to deduce stellar masses with a median value of $5 \times 10^9 M_{\odot}$, a factor of about two lower than previous virial estimates. They concluded that $L^*$\ compact blue galaxies at intermediate redshifts are about an order of magnitude less massive than typical $L^*$\ galaxies today.

Another lensed galaxy seen at high redshift and studied in detail is MS~1512-cB58 with $z = 2.73$ \citep{ell96, pet00, tep00}. This is an  $L^*$\ Lyman-break galaxy with a clearly measured stellar SED characterized by a continuous star formation model with a Salpeter IMF, an age of about 20 Myr and a virial mass of $1.2 \times 10^{10} M_{\odot}$. This has a metallicity of about 1/4 $Z_{\odot}$, a few times higher than typical damped \lya\ systems at the same redshift. In comparison, the Lynx arc is much younger, has a considerably lower metallicity and is less massive. The intrinsic, unlensed \hb\ luminosity of MS~1512-cB58 is some five times lower than that of the Lynx arc and it does not exhibit the strong ultraviolet emission spectrum which results from the latter's very hot ionizing stars.

\section{CONCLUSIONS}

The Lynx arc was discovered during the spectroscopic follow-up of the ROSAT Deep Cluster Survey. The fortuitous location of an \hii\ galaxy near a fold caustic in the source plane of a complex cluster environment at a redshift of a half has given us the opportunity to study the properties of an intrinsically luminous starforming region is some detail at a redshift of 3.357. 

From observations of the emission line spectrum from \lya\ to \neb\ we
show that the \hii\ region is characterized by very hot ionizing stars
with a temperature of $T_{\rm BB} \simeq 80,000$K and a high ionization parameter of $\log U \simeq -1$. While the nebular metallicity is not dramatically low,  with $Z/Z_{\odot} \sim 0.05$, there is indirect evidence that the ionizing stellar cluster is considerably more metal-poor. The low velocity dispersion of the nebula and the high luminosity imply a very low mass-to-light ratio for the part of the \hii\ galaxy that is strongly magnified. This conclusion is supported by the absence of any detectable stellar continuum in the spectrum longward of \lya: the observed flux in the restframe UV being satisfactorily explained as pure nebular continuum with no dust reddening. An unreddened young stellar cluster of $10^7 M_{\odot}$\ with a normal IMF would be detected in our data.

The outflowing wind seen as absorption components in both \lya\ and \civ\  is highly ionized, having similar column densities of $\approx10^{15}$~\cmmt\ in both H$^0$ and in C$^{3+}$. This is in contrast to local \hii\ galaxies which typically have H$^0$\ column densities closer to $10^{20}$~\cmmt\ \citep{kun98}.

From our photoionization modelling of the nebula, using scaled solar abundances, the most outstanding discrepancy amongst the observed and predicted emission line intensities is the \siiii\ doublet at $\lambda\lambda$1883, 1892. The substantial overabundance of silicon suggested by this observation may be a nucleosynthetic signature of past pair instability supernov\ae\ in a Population~III cluster \citep{heg02}. These supernov\ae\ result in the total disruption of stars in the mass range $140 - 260 M_{\odot}$ and produce an excess of elements with odd nuclear charge relative to the even nuclei.

It appears that we are witnessing a very early phase of star formation in a dwarf galaxy which initially had sufficiently low metallicity to allow the formation of many high mass stars producing an effective ionizing color temperature much higher than found in such objects in the local universe. The magnification-corrected ionizing luminosity of $Q_\mathrm{ion} \simeq 1.6 \times 10^{55}$~ph s$^{-1}$\ implies the presence of some 10$^6$\ hot O stars, while the low gas velocity dispersion and the absence of any detected stellar continuum longward of \lya\ demands a low mass-to-light ratio. The nebula has a high ionization parameter and there is an outflowing wind, ranging in velocity from $\sim 10$\ to $\sim 500$~\kms, which is much more highly ionized than winds in local \hii\ galaxies.

We are fortunate to have found such an object lensed by an intervening cluster of galaxies. Its intrinsic luminosity is, however, sufficiently high that astrophysically meaningful observations could be obtained of  similar objects that are unmagnified. The Lynx arc represents a galaxy that is considerably less massive than a typical Lyman break galaxy at a similar redshift \citep{pet01}\ but one that is in a very short-lived luminous phase lasting as little as a few million years and powered by a top-heavy IMF containing many stars with temperatures approaching $10^5$ K.

\acknowledgments Acknowledgments: This paper resulted from the
serendipitous discovery of the Lynx arc in the ROSAT Deep Cluster
Survey. We thank all the collaborators in that project who contributed
information and observational data.  We had useful discussions with
Luc Binette, Andrew Robinson, Stephane Charlot, and Daniel Schaerer.
We thank the Gemini/NIRSPEC service observing team for carrying out
the infrared spectroscopy at the Keck telescope with enthusiasm and
efficiency.

\clearpage

\begin{deluxetable}{lccccc}
\tablecaption{Emission line fluxes, intensities relative to \hb\ and,
  where measured from the ESI data, the instrument-corrected velocity
  dispersions of the narrowest component. The fluxes from the
  different observations have been scaled to represent the sum of the
  two arc components (A+B).  The predicted intensities of important
  lines from our fiducial photoionization model discussed in
  Section~\ref{photoionization} are given in the penultimate
  column. This model has: $T_{\rm BB} = 80,000$~K; $\log U = -1.0$ and $Z/Z_{\odot} = 0.05$. \label{line-fluxes}}
\tablewidth{0pt}
\tablehead{
\colhead{Line} & \colhead{$\lambda_{\rm rest}$}   & \colhead{$F_{\rm line}$}   &
\colhead{$I_{H_\beta=1}$}  &\colhead{$I_{H_\beta=1}$} & \colhead{$\sigma_{\rm line}$} \\
& \colhead{\AA} & \colhead{$10^{-16}$~\ecs} &\colhead{obs} & \colhead{model} & \colhead{\kms} }
\startdata
\ovi    & 1035  &       &     & 0.0002\\
\lya    & 1215  & 11.2  &3.11 & &\\
\lya$_{corr}$ && 38.9& 10.8 & 28.2& 150\\
\nv& 1240& $\leq$0.32& $\leq$0.09 & 0.02\\
\siiv  & 1397   & $\leq$0.32       & $\leq$0.09 & 0.05\\
\niv    & 1483  & 0.62  & 0.17 &\\
\niv& 1487& 0.88& 0.25 & 0.38\\
\civ& 1548+51& 2.32& 0.64& \\
\civ$_{corr}$& & 13.15& 3.65 & 7.11 & 80\\
\heii& 1640& 0.38& 0.11 & 0.25\\
\oiii& 1661& 0.66& 0.18 & \\
\oiii& 1666& 1.36& 0.38 & 0.94 & 35\\
\niii& 1750& 0.64& 0.18 & 0.05\\
\siiii& 1883& 0.32& 0.09 &0.0021\\
\siiii& 1892& 0.21& 0.06 &0.0016\\
\ciii& 1907&1.26& 0.35 & 1.6 & 30\\
\ciii& 1909& 0.87& 0.24& \\
\cii    &  2326 &       &       & 0.003 \\
\neiv&  2424    &       &       & 0.06  \\
\mgii&  2800    &       &       & 0.02 \\
\tableline
\oii& 3728& $\leq$0.9& $\leq$0.25 & 0.02\\
\neiii  & 3869  & 2.50  & 0.69 & 0.69\\
\neiii& 3969& $\leq$0.8& $\leq$0.22 &0.20\\
\neb & 4363     &       &       & 0.28 \\
\heii   & 4686  & $\leq$0.8        & $\leq$0.22 & 0.03\\
\hb& 4861& 3.60& 1.00 & 1.00\\
\neb& 4959& 9.30& 2.58 &\\
\neb& 5007& 27.00& 7.50 & 7.91\\
\hei   &  5876  &       &       & 0.09 \\
\oi    &  6300  &       &       & 0.0005 \\
\ha    &  6563  &       &       & 2.95 \\
\nii    &  6583 &       &       & 0.003 \\
\sii   &  6731  &       &       & 0.0007 \\

 \enddata
\tablecomments{The second entries for \lya\ and for \civ\ are values corrected for absorption. The rule below \ciii\ represents the division between optical and NIR spectroscopy. Above this line, the statistical flux errors ($1\sigma$) are approximately $1\times10^{-17}$~\ecs; below the line, they are approximately $3\times10^{-17}$~\ecs. Model intensities computed for multiplets, where shown as a single number, represent the sum of the components. It should be remembered that our data only extend to a restframe wavelength of 5400\AA.}
\end{deluxetable}

\end{document}